\PassOptionsToPackage{numbers}{natbib}
\documentclass[conference]{IEEEtran}
\IEEEoverridecommandlockouts
\usepackage{cite}
\usepackage{amsmath,amssymb,amsfonts}
\usepackage{algorithmic}
\usepackage{graphicx}
\usepackage{textcomp}
\usepackage{stfloats}
\usepackage{float}
\usepackage{xcolor}
\usepackage{booktabs}
\usepackage{multirow}
\usepackage{url}
\usepackage{natbib}
\usepackage{amsmath,amssymb,bm}
\usepackage{tikz}
\usetikzlibrary{calc,positioning,arrows.meta,backgrounds,fit}
\usepackage{booktabs}
\usepackage{xcolor}
\usepackage{tikz}
\usetikzlibrary{arrows.meta, positioning, fit, backgrounds}

\usetikzlibrary{shapes,arrows,positioning,calc}
\usepackage{amsmath}
\usepackage{amssymb}
\usepackage{caption}
\usepackage{cuted}  
\usepackage{placeins}

\def\BibTeX{{\rm B\kern-.05em{\sc i\kern-.025em b}\kern-.08em
    T\kern-.1667em\lower.7ex\hbox{E}\kern-.125emX}}
\begin{document}

\title{ORCA — Online Regime Correlation Analyzer
}

\author{
  \IEEEauthorblockN{Boris Kriuk}
  \IEEEauthorblockA{\textit{Department of Computer Science \& Engineering} \\
    \textit{Hong Kong University of Science and Technology}\\
    Clear Water Bay, Hong Kong \\
    bkriuk@connect.ust.hk}
  \and
  \IEEEauthorblockN{Fedor Kriuk}
  \IEEEauthorblockA{\textit{Faculty of Engineering \& Information Technology} \\
    \textit{University of Technology Sydney}\\
    Sydney, New South Wales, Australia \\
    fedor.kriuk@student.uts.edu.au}
}

\maketitle

\begin{abstract}
Standard risk models reduce the rich dependence fabric of
financial markets to scalar volatility estimates, discarding
the topological information encoded in cross-asset correlation
networks. We present ORCA (\textit{Online Regime Correlation
Analyzer}), an end-to-end framework that fuses spectral graph
theory, random matrix theory, and supervised machine learning
to deliver calibrated, continuously updated probability
estimates for both rally and crash events over a ten-day
forward horizon. ORCA constructs rolling correlation matrices
from 24 diversified exchange-traded instruments using three
parallel estimators operating at different time scales, and
extracts 127 spectral features---including absorption ratios,
eigenvalue entropy, effective rank, spectral gap, eigenvector
concentration metrics, and graph-topological descriptors at
multiple correlation thresholds---which are concatenated with
79 traditional price-derived indicators to form a
206-dimensional feature vector. A depth-limited Random Forest
with balanced sub-sample weighting is evaluated under a strict
eight-fold walk-forward protocol with ten-day anti-leakage
gaps spanning fifteen years of daily US market data. ORCA
achieves a Balanced Crisis Detection AUC
(\(\text{BCD-AUC} = \sqrt{\text{AUC}_{\text{rally}} \times
\text{AUC}_{\text{crash}}}\)) of 0.741, ranking first against
all baselines. Ablation studies confirm that spectral features
contribute +10.3 percentage points of AUC for crash detection
and +5.2 for rally detection relative to traditional features
alone, with SHAP analysis revealing that graph-topological
descriptors---clustering coefficient, edge density, and
dominant-eigenvalue percentile rank---constitute the three
most important crash predictors. A backtested ensemble
walk-forward strategy that maps the joint rally--crash signal
space to dynamic equity exposure with risk-on/risk-off
rotation achieves a Sharpe ratio of 1.13, a CAGR of 15.6\%,
and a maximum drawdown of only \(-7.5\%\), compared to 3.7\%
CAGR and \(-33.7\%\) drawdown for a buy-and-hold benchmark.
\end{abstract}


\section{Introduction}

Financial crises are characterised by a distinctive structural
signature: assets that are normally weakly correlated begin to
move in lockstep, compressing the eigenvalue spectrum of the
cross-asset correlation matrix well before headline volatility
indicators react. Standard risk models such as GARCH-family
estimators~\cite{engle1982autoregressive,bollerslev1986generalized} and realised-variance measures~\cite{andersen2003modeling,corsi2009simple} are inherently
reactive---they quantify the amplitude of price fluctuations
after a shock has materialised. By contrast, the
spectral properties of correlation networks shift
during the formation of a crisis: the dominant
eigenvalue absorbs an increasing share of total variance as
herding behaviour intensifies~\cite{kritzman2011principal,laloux1999noise,plerou2002random}, the spectral gap between the
first and second eigenvalues narrows, and the effective rank
of the correlation matrix collapses, signalling that
diversification is evaporating. Despite decades of evidence
for this phenomenon, most practitioner risk systems still
reduce the rich, high-dimensional dependence fabric of the
market to a single scalar---for example the VIX or a rolling
volatility estimate---discarding precisely the topological
information that could serve as an early warning.

This paper presents ORCA (\textit{Online Regime
Correlation Analyzer}), an end-to-end framework that fuses
spectral graph theory, random matrix theory, and supervised
machine learning to deliver calibrated, continuously updated
probability estimates for both rally and crash events over a
ten-day forward horizon. ORCA operates on a diversified
universe of 24 exchange-traded instruments spanning broad
equities, sector ETFs, international markets, fixed income,
commodities, and currencies. At each time step the system
constructs rolling correlation matrices using three parallel
estimators---a 60-day window capturing short-term regime
dynamics, a 120-day window for medium-term structure, and an
exponentially weighted estimator with a 30-day half-life for
adaptive smoothing---and extracts 127 spectral features from
their eigendecompositions. These features include absorption
ratios at multiple ranks (\(k \in \{1,3,5\}\)), eigenvalue
entropy and the derived effective rank, the
Marchenko--Pastur excess~\cite{marchenko1967distribution,bouchaud2009financial} measuring the distance from the
random-matrix null, the spectral gap as a herding indicator,
the condition number as a fragility proxy, eigenvector
concentration metrics such as the Herfindahl index of the
first-component loadings, graph-topological descriptors
(edge density, clustering coefficient, and degree
centralisation computed at correlation thresholds of 0.3,
0.5, and 0.7), and multi-horizon dynamics capturing the rate
of change, $z$-scores, and acceleration of all key spectral
quantities over 5-, 10-, and 20-day lookbacks. The spectral
features are augmented with 79 traditional price-derived
indicators---multi-horizon returns, realised and GARCH
volatility, downside semi-deviation, maximum intra-window
loss, momentum ratios, RSI, drawdown depth, higher moments
(skewness, kurtosis), volatility-of-volatility, and
cross-asset return dispersion---yielding a combined feature
vector of 206 dimensions.

The unified feature set is ingested by a Random Forest
classifier~\cite{breiman2001random} with 200 trees, a maximum depth of six, and
balanced sub-sample class weighting, pre-processed via a
RobustScaler. The model is evaluated under a strict
walk-forward protocol: eight expanding-window folds span
fifteen years of daily US market data, each with a three-year
training window, a six-month test window, and a ten-day gap
between them to prevent look-ahead contamination. All reported
metrics are purely out-of-sample. ORCA produces two outputs
per time step: the probability that the S\&P\,500 appreciates
by more than 3\% within ten trading days (rally detection,
base rate \(\approx 7.7\%\)), and the probability that the
index suffers an intra-window drawdown exceeding 7\% (crash
detection, base rate \(\approx 0.9\%\)). We evaluate both
tasks jointly via the Balanced Crisis Detection AUC
(\(\text{BCD-AUC} = \sqrt{\text{AUC}_{\text{rally}} \times
\text{AUC}_{\text{crash}}}\)), which penalises models that
are strong in only one direction.

On walk-forward evaluation ORCA achieves a BCD-AUC of
0.741, ranking first against competitive
baselines. Ablation studies isolate the
contribution of the spectral feature family at +5.2\%~AUC
for rally detection and +10.3\%~AUC for crash detection
relative to traditional features alone, confirming that
correlation network structure encodes crisis precursors
absent from price-based indicators. A backtested ensemble
walk-forward strategy that maps the two-dimensional signal
space of rally and crash ranks to a continuous equity
exposure---incorporating risk-on/risk-off rotation into a
defensive portfolio of Gold, intermediate Treasuries, and the
US dollar---translates these signals into a Sharpe ratio of
1.13, a CAGR of 15.6\%, and a maximum drawdown of
only \(-7.5\%\), compared to 3.7\% CAGR and \(-33.7\%\)
drawdown for a buy-and-hold benchmark, while operating at an
average leverage of just 0.39\(\times\).

The principal contributions of this work are threefold:

\begin{enumerate}
  \item Spectral--traditional feature fusion for
  balanced crisis detection. We formalise the integration of
  127 eigenvalue-derived, eigenvector-derived, and
  graph-topological features extracted from three parallel
  correlation estimators with 79 classical technical
  indicators, and demonstrate through ablation that the
  spectral family contributes an additional +10.3\%~AUC for
  crash detection and +5.2\%~AUC for rally detection---gains
  that are especially pronounced for the rarest and most
  economically consequential tail events.

  \item Walk-forward evaluation with the BCD-AUC
  metric. We design an eight-fold temporal cross-validation
  protocol with strict ten-day anti-leakage gaps and propose
  the BCD-AUC as a single summary statistic that rewards
  models effective in both the rally and crash directions,
  addressing the common pitfall of one-sided optimisation in
  crisis prediction literature. ORCA ranks first on this
  metric against all baselines evaluated.

  \item A deployable, auto-refreshing monitoring
  system with three-dimensional correlation-network
  visualisation. We implement the entire pipeline as a
  self-contained server that ingests live market data,
  re-estimates the full model every twelve hours, and serves
  an interactive dashboard in which the 24-asset
  correlation graph is rendered as a force-directed network
  with eigenvector-centrality-scaled nodes and
  correlation-coloured edges, enabling real-time visual
  monitoring of dependence topology shifts that would be
  invisible in conventional tabular displays.
\end{enumerate}

\section{Related Work}\label{sec:related}

The detection and prediction of financial crises has attracted
sustained attention across quantitative finance, econometrics,
and more recently machine learning. The literature most
relevant to ORCA spans four broad areas: volatility modelling
and forecasting, spectral analysis of financial correlation
matrices, network-based approaches to systemic risk, and
machine learning for tail-event prediction.

Volatility-based crisis detection has its roots in the
autoregressive conditional heteroskedasticity family of
models, beginning with the original ARCH specification~\cite{engle1982autoregressive} and
its generalised extension GARCH~\cite{bollerslev1986generalized}, which model time-varying
variance as a function of past squared returns and past
conditional variances. These models and their many variants,
including exponential GARCH, threshold GARCH, and realised
GARCH, remain the dominant paradigm for risk measurement in
practice. The heterogeneous autoregressive model of realised
volatility~\cite{corsi2009simple,andersen2003modeling} decomposes volatility into daily, weekly, and
monthly components and has proven surprisingly effective as a
simple linear benchmark. A common limitation of all such
approaches is that they are fundamentally univariate and
reactive: they measure the amplitude of price fluctuations
after a shock has arrived rather than detecting the structural
dependence shifts that precede it.

Random matrix theory provides the theoretical foundation for
distinguishing genuine correlation structure from estimation
noise in large-dimensional covariance matrices~\cite{laloux1999noise,plerou1999universal,plerou2002random,bouchaud2009financial,ledoit2004honey}. The
Marchenko--Pastur distribution~\cite{marchenko1967distribution} characterises the bulk
eigenvalue density of a purely random correlation matrix,
allowing researchers to identify eigenvalues that exceed the
noise threshold as carriers of true economic signal. The
absorption ratio, defined as the fraction of total variance
explained by a fixed number of top eigenvalues, was proposed
as a measure of systemic risk~\cite{kritzman2011principal,kritzman2010skulls,bisias2012survey} on the grounds that rising
absorption indicates increasing market fragility. Empirical
studies have documented that the absorption ratio tends to
increase in the months preceding major drawdowns and that its
rate of change can serve as a timing indicator. ORCA extends
this line of work by extracting a much richer set of spectral
descriptors, including eigenvalue entropy, effective rank,
spectral gap, condition number, and eigenvector concentration,
and by combining features from three parallel correlation
estimators operating at different time scales.

Network-based approaches to systemic risk model the financial
system as a graph in which nodes represent assets or
institutions and edges encode pairwise dependence, typically
measured by correlation, partial correlation, or mutual
information~\cite{billio2012econometric,preis2012quantifying,kenett2012quantifying}. Minimum spanning trees and planar maximally
filtered graphs~\cite{mantegna1999hierarchical,tumminello2005tool,bonanno2003topology,onnela2003dynamic,onnela2003dynamics,pozzi2013spread} have been used to track the evolving topology
of equity markets, revealing that network structure becomes
more centralised and less modular during stress episodes.
Granger-causality networks and transfer-entropy graphs~\cite{diebold2012better,diebold2014network} have
been employed to study directional spillovers across markets
and sectors. More recently, graph neural networks have been
applied to financial contagion modelling, learning node and
edge representations that capture higher-order relational
structure. ORCA draws on graph-topological features such as
edge density, clustering coefficient, and degree
centralisation computed at multiple correlation thresholds,
but deliberately avoids deep graph neural architectures in
favour of interpretable ensemble tree models, which are better
suited to the small-sample, non-stationary setting of crisis
prediction.

Machine learning methods for tail-event forecasting have
explored gradient-boosted trees~\cite{chen2016xgboost}, support vector machines,
long short-term memory networks~\cite{hochreiter1997long}, and attention-based
architectures for predicting market crashes, drawdowns, and
volatility spikes~\cite{chatzis2018forecasting,kriuk2025deepsupp,alkhamov2025equity, shi2020change}. A recurring finding is that ensemble tree
methods such as random forests~\cite{breiman2001random} and gradient-boosted machines~\cite{chen2016xgboost,kriuk2025morphboost}
offer a favourable bias-variance trade-off for tabular
financial data, particularly when combined with careful
temporal cross-validation to avoid look-ahead bias. Class
imbalance, which is acute for extreme events occurring on
fewer than one percent of trading days, is typically addressed
through resampling, cost-sensitive learning, or threshold
optimisation. ORCA adopts balanced sub-sample class weighting
within a depth-limited random forest and evaluates performance
using threshold-free metrics, specifically AUC and average
precision, to avoid artefacts introduced by arbitrary
probability cut-offs.

Despite the breadth of prior work, no existing framework
simultaneously extracts multi-scale spectral and topological
features from dynamic correlation networks, fuses them with
traditional volatility and momentum indicators, evaluates
both rally and crash detection under a unified balanced
metric, and deploys the resulting model as a continuously
refreshing monitoring system with real-time network
visualisation. ORCA aims to fill this gap.

\section{Methodology}\label{sec:method}

This section describes the ORCA pipeline in full: the
multi-asset universe and correlation estimation scheme, the
spectral and topological feature extraction, the traditional
feature baseline, the supervised classification model, the
walk-forward evaluation protocol, and the regime-to-exposure
mapping used in the backtested strategy.

\subsection{Data and correlation estimation}

ORCA ingests daily adjusted close prices for a universe of 24
exchange-traded instruments spanning six asset classes:
broad equities (S\&P\,500, Nasdaq\,100, Russell\,2000),
US sector ETFs (Financials, Energy, Technology, Healthcare,
Utilities, Consumer Staples, Consumer Discretionary,
Industrials, Materials), international equities (Developed
International, Emerging Markets, Europe, Japan), fixed income
(Long Treasury, Intermediate Treasury, Investment-Grade
Corporate, High Yield), commodities (Gold, Oil), currencies
(US Dollar), and real estate (REITs). Prices are sourced via
the EODHD API with local disk caching and cover fifteen years
of history. After forward-filling gaps of up to five trading
days and dropping residual missing values, daily simple
returns are computed as
\(r_{i,t} = p_{i,t}/p_{i,t-1} - 1\).

At each date \(t\) we construct three parallel estimates of
the \(n \times n\) cross-asset correlation matrix, where
\(n = 24\). The first uses a 60-day trailing window of
returns, the second a 120-day window, and the third an
exponentially weighted covariance estimator with a half-life
of 30 days. For the exponential estimator the covariance
matrix \(\Sigma^{\mathrm{ewm}}_t\) is computed from the
exponentially weighted second moments and then converted to
correlation form by normalising each entry by the product of
the corresponding marginal standard deviations. All three
matrices are symmetrised, have their diagonals set to unity,
and are clipped to the interval \([-1, 1]\). The use of
multiple windows allows the model to distinguish short-term
regime shifts (60-day) from medium-term structural changes
(120-day) and to benefit from the adaptive smoothing
properties of exponential weighting.

\subsection{Spectral and topological features}

Given a correlation matrix
\(C_t \in \mathbb{R}^{n \times n}\), we compute its
eigendecomposition
\(C_t = V \Lambda V^{\top}\), where
\(\Lambda = \mathrm{diag}(\lambda_1, \ldots, \lambda_n)\)
with \(\lambda_1 \geq \lambda_2 \geq \cdots \geq \lambda_n
\geq 0\). The eigenvalues and eigenvectors give rise to the
following feature families.

The absorption ratio at rank \(k\) measures the fraction of
total variance captured by the top \(k\) eigenvalues:
\[
  \mathrm{AR}_k
  \;=\;
  \frac{\sum_{i=1}^{k} \lambda_i}
       {\sum_{i=1}^{n} \lambda_i}\,.
\]
We extract \(\mathrm{AR}_1\), \(\mathrm{AR}_3\), and
\(\mathrm{AR}_5\). A rising absorption ratio indicates that
variance is concentrating into fewer dimensions, a hallmark
of herding behaviour.

The eigenvalue entropy quantifies how uniformly variance is
distributed across the spectrum:
\[
  H
  \;=\;
  -\sum_{i=1}^{n} \tilde{\lambda}_i
    \ln \tilde{\lambda}_i\,,
  \qquad
  \tilde{\lambda}_i
  = \frac{\lambda_i}{\sum_{j=1}^{n} \lambda_j}\,.
\]
From the entropy we derive the effective rank as
\(\mathrm{ER} = \exp(H)\), which can be interpreted as the
number of independent risk factors driving the market. During
normal conditions \(\mathrm{ER}\) is close to \(n\); during
crises it collapses toward unity.

The spectral gap is defined as the ratio of the first to the
second eigenvalue,
\(\gamma = \lambda_1 / \lambda_2\). A large gap indicates
that a single market factor dominates, consistent with
panic-driven co-movement. We also record the condition number
\(\kappa = \lambda_1 / \lambda_n\), the Marchenko--Pastur
excess
\(\lambda_1 - (1 + \sqrt{n/T})^2\) where \(T\) is the
estimation window length, eigenvalue standard deviation,
skewness, and kurtosis.

From the first eigenvector
\(\mathbf{v}_1 = (v_{1,1}, \ldots, v_{1,n})^{\top}\)
we extract concentration metrics. The normalised absolute
loadings are
\(\bar{v}_{1,i} = |v_{1,i}| / \sum_j |v_{1,j}|\), from
which we compute the Herfindahl index
\(\mathrm{HHI} = \sum_i \bar{v}_{1,i}^{2}\), loading
entropy, maximum loading, and the dispersion between the
top-quartile and bottom-quartile loadings. High concentration
signals that the crisis factor loads unevenly on the asset
universe, while uniform loadings indicate broad market stress.

For the graph-topological features, we threshold the absolute
correlation matrix at levels \(\tau \in \{0.3, 0.5, 0.7\}\)
to obtain binary adjacency matrices
\(A^{(\tau)}_{ij} = \mathbf{1}[|C_{t,ij}| > \tau]\) with
zeros on the diagonal. At each threshold we compute edge
density, mean degree, degree standard deviation, maximum
degree, the number of isolated nodes, Freeman degree
centralisation, and the global clustering coefficient
\[
  \mathcal{C}^{(\tau)}
  \;=\;
  \frac
    {\sum_i \sum_{j<k} A^{(\tau)}_{ij}\, A^{(\tau)}_{jk}\, A^{(\tau)}_{ik}}
    {\sum_i \binom{d_i^{(\tau)}}{2}}\,,
\]
where \(d_i^{(\tau)} = \sum_j A^{(\tau)}_{ij}\) is the
degree of node \(i\). We additionally record aggregate
correlation statistics: mean, median, and maximum absolute
correlation, correlation standard deviation and skewness, and
the fraction of pairs exceeding 0.50 and 0.70.

To capture the dynamics of regime transitions, we compute for
each of eight key spectral quantities --- the dominant
eigenvalue, its variance share, the rank-one absorption
ratio, eigenvalue entropy, effective rank, mean absolute
correlation, edge density at 0.50, and clustering coefficient
at 0.50 --- a set of temporal derivatives over horizons of 5,
10, and 20 trading days: percentage rate of change, absolute
difference, and a rolling z-score computed against a window of
twice the horizon length. We also record the acceleration
(second difference) and a 252-day trailing percentile rank
for each quantity. The idea yields a dynamics feature set that
tracks not only the level but also the speed and acceleration
of spectral regime change.

All spectral, eigenvector, topological, and dynamics features
are extracted from each of the three correlation estimators,
producing a total of 127 features~\cite{kriuk2026poseidon}.

\begin{figure}[t]
\centering
\resizebox{\columnwidth}{!}{%
\begin{tikzpicture}[
    >=Stealth,
    node distance=0.8cm,
    block/.style={
        rectangle, draw, rounded corners=3pt,
        minimum width=2.8cm, minimum height=0.65cm,
        align=center, font=\scriptsize\sffamily
    },
    data/.style={block, fill=blue!12},
    proc/.style={block, fill=orange!12},
    feat/.style={block, fill=green!12},
    mdl/.style={block, fill=red!12},
    res/.style={block, fill=violet!12},
    strat/.style={block, fill=cyan!12},
    arr/.style={->, thick, black!70},
    darr/.style={->, dashed, black!50},
    glabel/.style={font=\tiny\sffamily\bfseries, text=black!50},
]

\node[data] (prices) {Daily Prices (24 assets)};

\node[data, below=0.6cm of prices] (returns) {Returns \(r_{i,t}\)};
\draw[arr] (prices) -- (returns);

\node[proc, below left=0.8cm and 1.6cm of returns, minimum width=1.8cm] (c60) {60d Corr};
\node[proc, below=0.8cm of returns, minimum width=1.8cm] (c120) {120d Corr};
\node[proc, below right=0.8cm and 1.6cm of returns, minimum width=1.8cm] (cewm) {EWM Corr};

\draw[arr] (returns) -- (c60);
\draw[arr] (returns) -- (c120);
\draw[arr] (returns) -- (cewm);

\node[feat, below=0.8cm of c60, minimum width=1.8cm] (eigf) {\scriptsize Eigenvalue\\[-1pt]\scriptsize feat.};
\node[feat, below=0.8cm of c120, minimum width=1.8cm] (vecf) {\scriptsize Eigenvector\\[-1pt]\scriptsize feat.};
\node[feat, below=0.8cm of cewm, minimum width=1.8cm] (topf) {\scriptsize Topology\\[-1pt]\scriptsize feat.};

\draw[arr] (c60) -- (eigf);
\draw[arr] (c120) -- (vecf);
\draw[arr] (cewm) -- (topf);
\draw[darr] (c60) -- (vecf);
\draw[darr] (c120) -- (eigf);
\draw[darr] (c120) -- (topf);

\node[feat, below=0.7cm of vecf, minimum width=2.2cm] (dyn) {Dynamics (ROC, z, accel)};
\draw[arr] (eigf) -- (dyn);
\draw[arr] (vecf) -- (dyn);
\draw[arr] (topf) -- (dyn);

\node[draw=green!50, dashed, rounded corners=4pt,
      fit=(eigf)(vecf)(topf)(dyn),
      inner xsep=0.2cm, inner ysep=0.15cm,
      label={[glabel]above left:{127 spectral feat.}}] (sbox) {};

\node[feat, right=0.5cm of topf, minimum width=1.8cm] (trad) {\scriptsize Traditional\\[-1pt]\scriptsize (79 feat.)};

\draw[arr] (returns.east) -| (trad.north);

\node[mdl, below=1.0cm of dyn, xshift=0.8cm] (fuse) {Concatenation (206 dim)};
\draw[arr] (dyn) -- (fuse);
\draw[arr] (trad) |- (fuse);

\node[mdl, below=0.6cm of fuse] (scaler) {RobustScaler};
\draw[arr] (fuse) -- (scaler);

\node[mdl, below=0.5cm of scaler] (rf) {Random Forest (200 trees, d=6)};
\draw[arr] (scaler) -- (rf);

\node[draw=red!40, dashed, rounded corners=4pt,
      fit=(scaler)(rf),
      inner sep=0.15cm,
      label={[glabel]right:{WF-CV (8 folds)}}] (wf) {};

\node[res, below left=0.8cm and 0.9cm of rf, minimum width=2.0cm] (rally) {\(\hat{p}_{\mathrm{rally}}\)};
\node[res, below right=0.8cm and 0.9cm of rf, minimum width=2.0cm] (crash) {\(\hat{p}_{\mathrm{crash}}\)};
\draw[arr] (rf) -- (rally);
\draw[arr] (rf) -- (crash);

\node[res, below=1.0cm of rf, yshift=-0.8cm] (rank) {126d Percentile Rank};
\draw[arr] (rally) -- (rank);
\draw[arr] (crash) -- (rank);

\node[strat, below left=0.8cm and 0.9cm of rank, minimum width=2.0cm] (regime) {\scriptsize Regime\\[-1pt]\scriptsize (5 states)};
\node[strat, below right=0.8cm and 0.9cm of rank, minimum width=2.0cm] (exposure) {\scriptsize Exposure\\[-1pt]\scriptsize \(w_t\!\in\![0,1.5]\)};
\draw[arr] (rank) -- (regime);
\draw[arr] (rank) -- (exposure);

\node[strat, below=1.0cm of rank, yshift=-0.8cm] (stgy) {RORO Strategy};
\draw[arr] (regime) -- (stgy);
\draw[arr] (exposure) -- (stgy);

\node[glabel, right=0.3cm of crash] {BCD-AUC};

\end{tikzpicture}
}%
\caption{ORCA architecture. Solid arrows show primary data flow; dashed arrows indicate cross-estimator feature sharing. WF-CV denotes walk-forward cross-validation.}
\label{fig:architecture}
\end{figure}
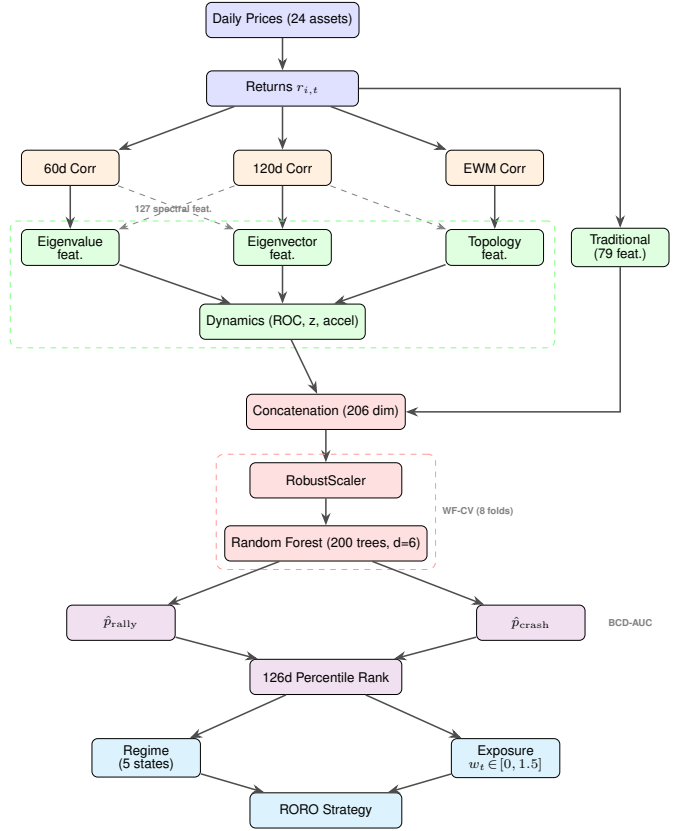

\subsection{Traditional features}

To provide a competitive baseline and to supply the model
with price-level information absent from the correlation
spectrum, we construct 79 traditional features from the
S\&P\,500 price series and the multi-asset return panel.
These include multi-horizon returns at 1, 5, 10, 20, and 60
days; rolling annualised volatility at 5, 10, 20, and 60
days; volatility ratios (5-to-20 day, 10-to-60 day); a
GARCH(1,1) conditional volatility estimate with parameters
\(\alpha = 0.1\), \(\beta = 0.85\); downside semi-deviation
over 20 days; maximum single-day loss over 5 and 20 days;
price-to-moving-average ratios at 10, 20, and 50 days; a
14-day RSI; drawdown from rolling 20-day and 60-day highs;
rolling skewness and kurtosis at 20 and 60 days;
volatility-of-volatility (the 20-day standard deviation of
5-day rolling volatility); and cross-asset return dispersion,
defined as the cross-sectional standard deviation of daily
returns across all 24 instruments. The 127 spectral features
and 79 traditional features are concatenated into a combined
vector of 206 dimensions.

\subsection{Classification model and walk-forward protocol}

The combined feature vector is pre-processed with a robust
scaler (centering on the median and scaling by the
inter-quartile range) and any residual non-finite values are
replaced with zero. The scaled features are fed to a Random
Forest classifier~\cite{breiman2001random,pedregosa2011scikit} with 200 trees, a maximum depth of six, a
minimum of 30 samples per leaf, a minimum of 60 samples per
split, and balanced sub-sample class weighting to address the
severe class imbalance inherent in rare-event prediction.

We define two binary prediction targets on the S\&P\,500
over a ten-day forward window. The rally target equals one
when the endpoint return exceeds three percent. The crash
target equals one when the maximum intra-window drawdown
from the entry price exceeds seven percent. Base rates are
approximately 7.7 percent and 0.9 percent respectively.

Evaluation follows a strict walk-forward protocol with eight
expanding-window folds. Each fold uses a three-year training
window, a ten-day gap to prevent information leakage from
overlapping return horizons, and a six-month out-of-sample
test window. Folds are laid out sequentially with no overlap
between test periods. All metrics reported in this paper are
computed exclusively on out-of-sample predictions aggregated
across folds.

We evaluate both tasks with threshold-free metrics to avoid
artefacts from arbitrary probability cut-offs: AUC-ROC and
average precision. For precision, recall, and F1 we select
the probability threshold that maximises F1 on the
out-of-sample predictions, clipped to the interval
\([0.05,\, 0.95]\). To obtain a single summary statistic that
rewards models effective in both directions, we define the
Balanced Crisis Detection AUC:
\[
  \mathrm{BCD\text{-}AUC}
  \;=\;
  \sqrt{
    \mathrm{AUC}_{\mathrm{rally}}
    \;\times\;
    \mathrm{AUC}_{\mathrm{crash}}
  }\,.
\]
The geometric mean penalises models that are strong in only
one direction more severely than the arithmetic mean would,
which is desirable because a system that detects rallies but
misses crashes (or vice versa) is of limited practical value.

\subsection{Regime assignment}

The raw rally and crash probabilities produced by the model
are converted to 126-day rolling percentile ranks,
\(\hat{r}_t\) and \(\hat{c}_t\) respectively, to make
thresholds stationary across market regimes. A deterministic
rule maps the two-dimensional signal
\((\hat{r}_t, \hat{c}_t)\) to one of five regimes: Normal,
Rally, Caution, Euphoria, and Crisis. The same pair is mapped
to a continuous equity exposure multiplier
\(w_t \in [0, 1.5]\) via a piecewise function calibrated on
conditional return analysis. When \(\hat{r}_t \geq 0.90\) or
\(\hat{c}_t \geq 0.60\) the exposure is set to zero, as the
former condition corresponds to rally overconfidence (which
historically precedes negative returns) and the latter to
elevated crash risk. When
\(0.78 \leq \hat{r}_t < 0.90\) and \(\hat{c}_t < 0.40\) the
exposure reaches its maximum of \(1.5\times\). Intermediate
zones receive exposures of 0.3, 0.7, 1.0, or 1.2 depending
on the crash rank.


\section{Experiments}
\label{sec:experiments}

\subsection{Experimental Setup}

We evaluate ORCA on 15 years of daily US market data (2009--2024) spanning 24 diversified assets across six categories: broad equity (SPY, QQQ, IWM), sectors (XLF, XLE, XLK, XLV, XLU, XLP, XLY, XLI, XLB, XLRE), international (EFA, EEM, VGK, EWJ), fixed income (TLT, IEF, LQD, HYG), commodities (GLD, USO), and currencies (UUP). The universe is intentionally heterogeneous: spectral features derive their predictive power from cross-asset correlation dynamics, and a diverse universe ensures the correlation matrix captures genuine structural shifts rather than sector-specific noise.

All data are sourced from the EODHD API using adjusted close prices. Returns are computed as simple daily percentage changes. Missing values are forward-filled up to five days; remaining gaps are dropped. The final aligned dataset contains approximately 3{,}750 trading days across all 24 assets.

\subsection{Prediction Tasks}

We define two complementary binary classification tasks designed to capture extreme tail events in both directions:

\textbf{Rally Detection.} The target is 1 if the S\&P~500 rises by more than 3\% at any point within the next 10 trading days, measured as the endpoint return \(r_{t+10} = P_{t+10}/P_t - 1 > 0.03\). The base rate is approximately 7.7\%, making it a moderately imbalanced problem.

\textbf{Crash Detection.} The target is 1 if the S\&P~500 experiences a maximum intra-window drawdown exceeding 7\% within the next 10 trading days, computed as \(\min_{k \in [1,10]} P_{t+k}/P_t - 1 < -0.07\). The base rate is approximately 0.9\%, making it a severely imbalanced rare-event detection problem. The drawdown formulation is deliberately more conservative than an endpoint return measure: it captures intra-period stress that an end-of-window snapshot would miss.

These two tasks are evaluated jointly through the Balanced Crisis Detection AUC (BCD-AUC), defined as the geometric mean of the per-task AUC-ROC scores:
\[
\text{BCD-AUC} = \sqrt{\text{AUC}_{\text{rally}} \times \text{AUC}_{\text{crash}}}
\]
Such metric penalises models that excel in one direction at the expense of the other, rewarding balanced detection capability. 

\subsection{Walk-Forward Validation Protocol}

Temporal data prohibits standard cross-validation. We employ a strict walk-forward protocol with anti-leakage guarantees:

\begin{enumerate}
    \item \textbf{Training window:} 3 years.
    \item \textbf{Gap:} 10 trading days between the end of training and start of testing, preventing any target leakage from the 10-day prediction horizon.
    \item \textbf{Test window:} 6 months .
    \item \textbf{Folds:} 8 non-overlapping test periods .
\end{enumerate}

All reported metrics are computed exclusively on out-of-sample test folds. No hyperparameter tuning is performed on test data; the Random Forest configuration (200 trees, max depth 6, minimum 30 samples per leaf, balanced\_subsample class weighting) is fixed across all folds and tasks.

\subsection{Baselines}

We compare ORCA against five competitive baselines spanning the spectrum from simple heuristics to established financial econometric models~\cite{corsi2009simple,kritzman2011principal,kritzman2010skulls,bollerslev1986generalized}.

\subsection{Classification Results}

Table~\ref{tab:classification} presents the walk-forward out-of-sample results for all six models across both tasks.

\begin{table}[t]
\centering
\caption{Walk-forward out-of-sample classification performance. BCD-AUC = \(\sqrt{\text{Rally} \times \text{Crash}}\). Bold indicates best per column.}
\label{tab:classification}
\renewcommand{\arraystretch}{1.15}
\footnotesize
\begin{tabular}{@{}lcccc@{}}
\toprule
\textbf{Model} & \textbf{Rally} & \textbf{Crash} & \textbf{BCD-AUC} & \textbf{\#Feat} \\
\midrule
\textbf{ORCA (Ours)}     & \textbf{0.772} & 0.711          & \textbf{0.741} & 206 \\
HAR-RV                    & 0.696          & \textbf{0.763} & 0.729          & 4   \\
SMA Volatility            & 0.682          & 0.671          & 0.676          & 9   \\
Traditional RF            & 0.736          & 0.608          & 0.669          & 27  \\
Spectral Only RF          & 0.620          & 0.666          & 0.643          & 179 \\
Turbulence                & 0.647          & 0.565          & 0.605          & 5   \\
\bottomrule
\end{tabular}
\end{table}

ORCA achieves the highest BCD-AUC of 0.741, ranking first overall. Several patterns merit discussion.

First, ORCA dominates on Rally AUC (0.772) by a substantial margin over all baselines, with the nearest competitor being Traditional RF at 0.736. Such observation suggests that the combination of spectral and traditional features captures upside momentum signals that neither feature set achieves alone.

Second, HAR-RV achieves the highest Crash AUC (0.763), outperforming ORCA's 0.711 on this individual task. However, HAR-RV's Rally AUC of 0.696 is significantly weaker, resulting in a lower BCD-AUC of 0.729. The pattern illustrates the value of the geometric mean metric: HAR-RV's crash detection strength comes at the cost of rally detection, while ORCA maintains high performance on both.

Third, the Turbulence baseline performs poorly on both tasks (BCD-AUC 0.605), despite being a widely cited crisis indicator. It is consistent with the understanding that Mahalanobis distance is a coincident rather than leading indicator: it peaks during crises rather than before them.

\subsection{Ablation Study}

Table~\ref{tab:ablation} isolates the contribution of spectral features
by comparing three feature configurations under identical model and
walk-forward validation settings.

\begin{table}[t]
\centering
\caption{Ablation study: spectral feature contribution.
$\Delta$ is measured relative to the Traditional-only baseline.}
\label{tab:ablation}
\renewcommand{\arraystretch}{1.15}
\footnotesize
\begin{tabular}{@{}lccc@{}}
\toprule
\textbf{Feature Set} & \textbf{Rally AUC} & \textbf{Crash AUC}
  & $\boldsymbol{\Delta}$ \textbf{vs Trad.} \\
\midrule
Traditional only  & 0.736 & 0.608 & baseline \\
Spectral only     & 0.620 & 0.666 & $-$0.116\,/\,+0.058 \\
Combined (ORCA)   & 0.772 & 0.711 & +0.036\,/\,+0.103 \\
\bottomrule
\end{tabular}
\end{table}

The spectral contribution is asymmetric. For crash detection, adding
spectral features yields a +10.3~pp AUC improvement over
traditional features alone ($0.608 \to 0.711$), the largest
single-source gain in our experiments. Spectral features alone already
surpass traditional features on this task ($0.666$ vs.\ $0.608$),
confirming the central hypothesis: correlation network structure encodes
crisis precursors absent from univariate price-based features.

For rally detection, spectral features in isolation underperform
($0.620$ vs.\ $0.736$), yet the combination yields the best score
($0.772$, a +3.6~pp lift). This complementarity demonstrates that
spectral features supply orthogonal regime-level context rather than
duplicating momentum or volatility signals.

\begin{figure}[t]
\centering
\includegraphics[width=\columnwidth]{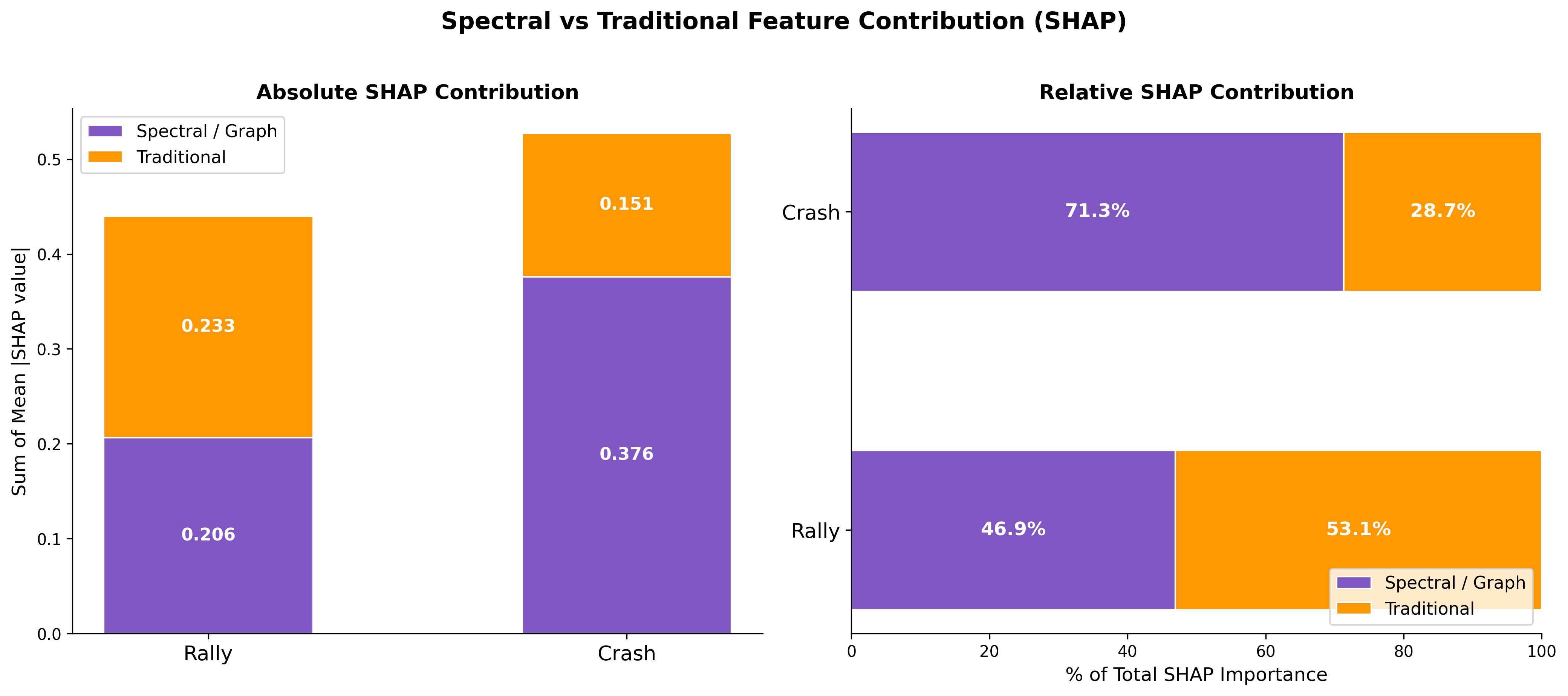}
\caption{Aggregated SHAP importance by feature family.}
\label{fig:shap_contribution}
\end{figure}

Fig.~\ref{fig:shap_contribution} quantifies this asymmetry through
aggregated SHAP importance. In absolute terms the crash model's total
SHAP mass is dominated by spectral/graph features ($0.376$ vs.\
$0.151$), translating to a 71.3\% relative share---nearly
three times the contribution of traditional features. The rally model is
more balanced (spectral 46.9\%, traditional 53.1\%), consistent with the
ablation table: traditional momentum and drawdown features carry the
rally signal while spectral features refine it at the margin.

\subsection{Feature Importance Analysis}

We examine individual feature importances and their nonlinear dependence
profiles via SHAP~\cite{lundberg2017unified}, computed on pooled
out-of-sample predictions across all eight walk-forward folds.

\begin{figure*}[t]
\centering
\includegraphics[width=\textwidth]{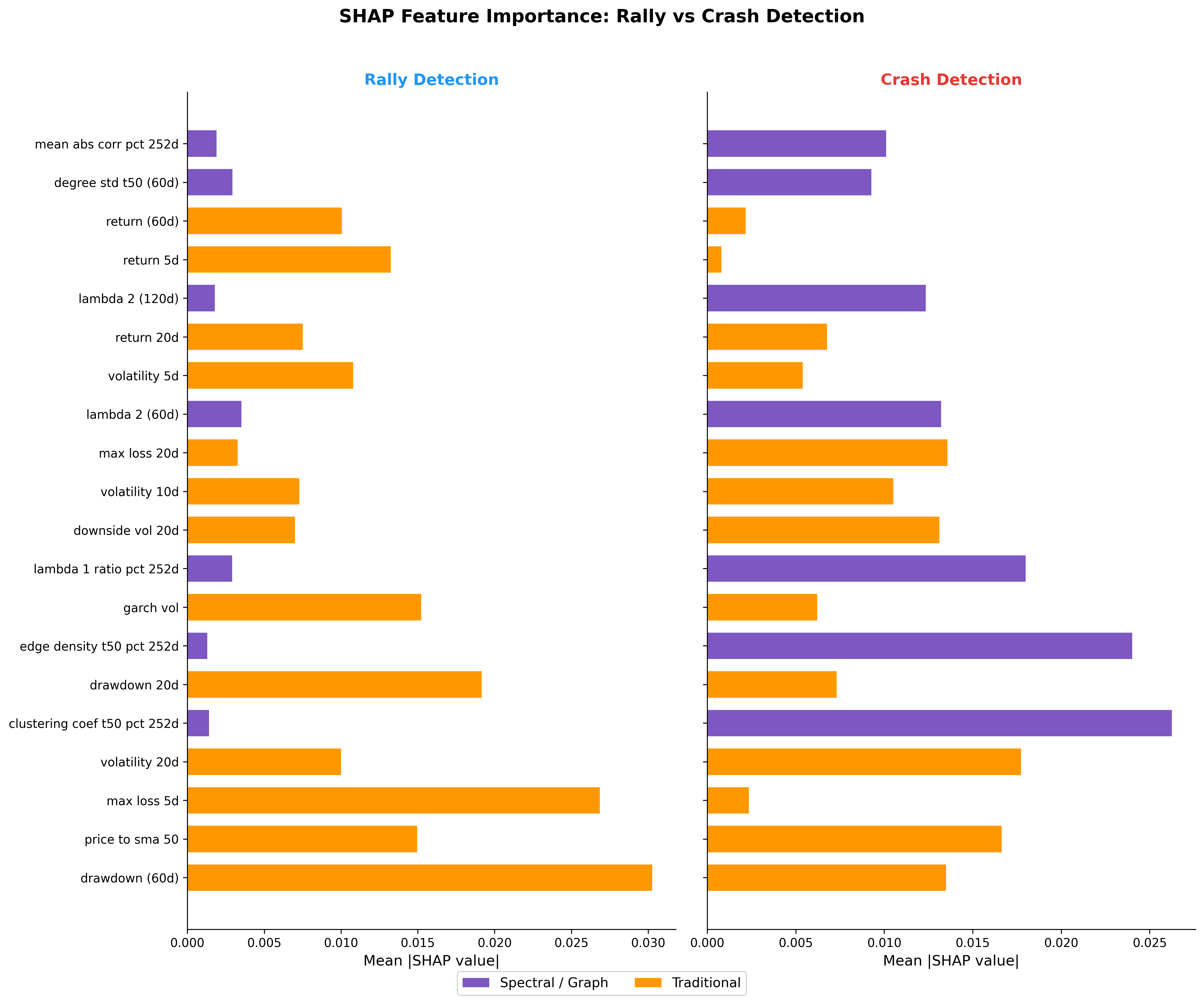}
\caption{SHAP feature importance for rally
(left) and crash (right) detection. The crash model is led by graph-topological features
while the rally model is led by traditional drawdown and momentum
features.}
\label{fig:shap_importance}
\end{figure*}

Fig.~\ref{fig:shap_importance} ranks the top 20 features by mean
absolute SHAP value for each task. For rally detection the four most
important features---\texttt{drawdown\,(60d)},
\texttt{price\_to\_sma\_50}, \texttt{max\_loss\_5d}, and
\texttt{drawdown\_20d}---are all price-based, capturing mean-reversion
after distress. Spectral features appear lower in the ranking,
contributing refinement rather than primary signal. For crash detection
the picture inverts: the top three features---\texttt{clustering\_coef\_
t50\_pct\_252d}, \texttt{edge\_density\_t50\_pct\_252d}, and
\texttt{lambda\_1\_ratio\_pct\_252d}---are all graph-topological,
tracking how tightly correlated the asset universe has become relative to
its recent history. Traditional features such as
\texttt{price\_to\_sma\_50} and \texttt{volatility\_20d} remain relevant
but occupy secondary ranks.

\begin{figure*}[p]
\centering
\includegraphics[width=\textwidth]{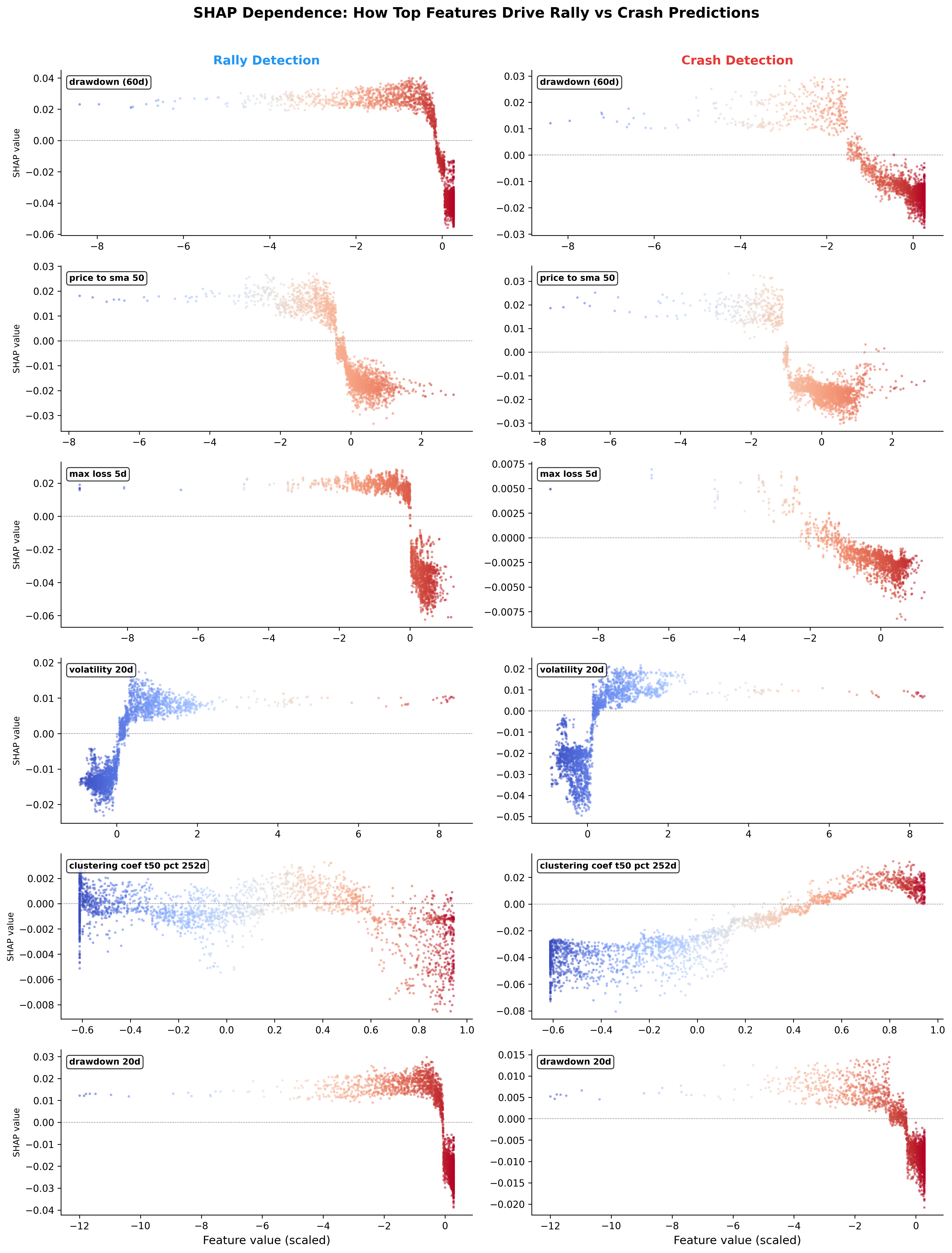}
\caption{SHAP dependence plots for the six most important features.
Left column: rally detection; right column: crash detection. The
$x$-axis is the scaled feature value; the $y$-axis is the SHAP
contribution.}
\label{fig:shap_dependence}
\end{figure*}

Fig.~\ref{fig:shap_dependence} presents SHAP dependence plots for the
six most influential features side-by-side for rally (left column) and
crash (right column). Each point is one out-of-sample observation;
colour runs from blue (low feature value) to red (high).

\textbf{Drawdown\,(60d).} For rallies the relationship is a hockey
stick: SHAP is near zero when the drawdown is shallow but swings
sharply positive as it deepens beyond $\approx -4\sigma$, capturing
mean-reversion after distress. For crashes the dependence inverts: deep
prior drawdowns produce negative SHAP, reflecting the empirical
regularity that severe crashes rarely compound without a bounce.

\textbf{Price-to-SMA\,50.} Prices well below the 50-day average
generate positive rally SHAP (mean-reversion) and mildly positive crash
SHAP (further distress risk). Prices well above the average suppress
rally SHAP, consistent with overextension.

\textbf{Max loss\,5d.} Extreme recent losses generate strongly negative
rally SHAP---acute stress suppresses immediate rebound predictions---but
only weakly affect crash SHAP, indicating the crash model relies more on
structural (spectral) than acute (price) stress signals.

\textbf{Volatility\,20d.} Both tasks show a two-regime structure. At low
volatility SHAP fans out widely (uninformative). At elevated volatility
($>4\sigma$) both models assign positive SHAP: vol spikes signal a tail
event, but the direction is resolved by other features.

\textbf{Clustering coefficient.} This is the most striking dependence.
For crash detection a sharp cliff appears: when the 252-day percentile
drops below $\approx -0.4$ (the correlation network becoming unusually
sparse), SHAP plummets to $-0.06$ to $-0.08$, strongly pushing toward
the crash class. A sudden collapse in clustering signals decorrelation
among formerly linked assets---a hallmark of liquidity withdrawal and
flight-to-quality dynamics preceding market dislocations. For rally
detection the same feature has near-zero SHAP across its range,
confirming its crash-specific informational content.

\textbf{Drawdown\,20d.} Mirrors the 60-day version with shorter memory.
Deep 20-day drawdowns yield positive rally SHAP, but for crash detection
the effect is attenuated relative to spectral features, reinforcing the
hierarchy: network topology matters more than short-term drawdowns for
anticipating systemic events.

Taken together the SHAP analysis provides a mechanistic explanation for
the ablation results: traditional drawdown and momentum features
dominate rally detection because rallies are price-driven mean-reversion
events. Crashes, by contrast, are preceded by structural shifts in the
correlation network---captured by clustering coefficients, edge
densities, and eigenvalue ratios---invisible to univariate price
statistics. ORCA succeeds because it fuses both signal families,
recreating balanced tail-event detection in either direction.


\section{Trading Strategy}
\label{sec:strategy}

\begin{figure}[!htbp]
\centering
\includegraphics[width=\columnwidth]{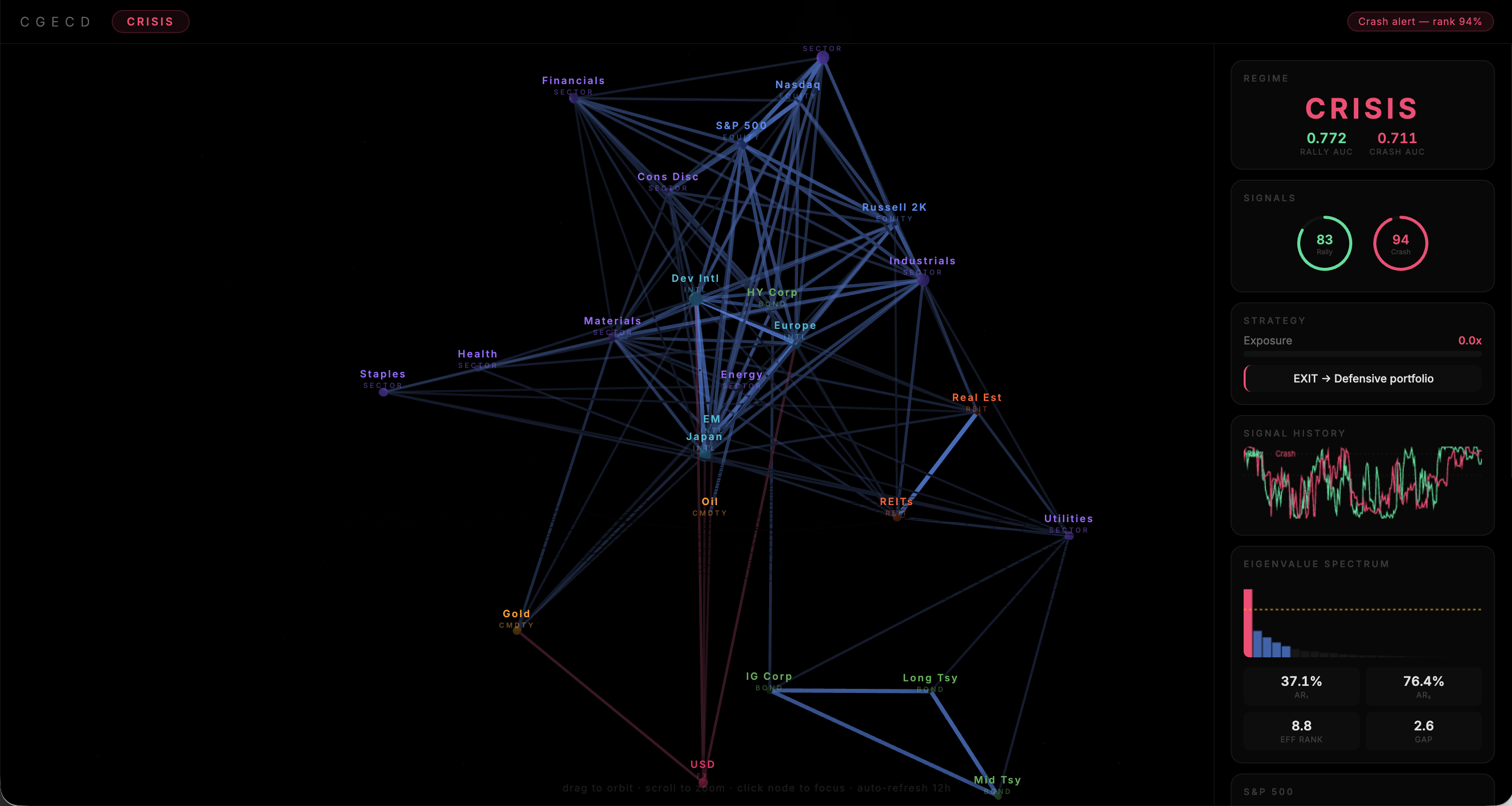}
\caption{Live demonstration of regime-dependent portfolio behaviour. High-exposure periods capture equity upside, while defensive periods rotate into safe-haven assets during market drawdowns. Can be found at https://orca.boriskriuk-powered.com/}
\label{fig:demo}
\end{figure}

To assess whether ORCA's classification signals translate into economically meaningful portfolio decisions, we design a single backtested trading strategy---Ensemble Walk-Forward Optimisation (Ensemble WFO)---that converts rally and crash probability estimates into dynamic equity exposure. The strategy achieves a Sharpe ratio of 1.13 with 15.6\% CAGR and only $-$7.5\% maximum drawdown, compared to the S\&P~500 buy-and-hold benchmark's Sharpe of 0.09 and $-$33.7\% maximum drawdown over the same period.

\subsection{Signal Processing}

The raw model outputs $\hat{p}_{\text{rally},t}$ and $\hat{p}_{\text{crash},t}$ are non-stationary: their distributional properties shift across walk-forward folds as market regimes evolve. To normalise these signals, we apply a rolling 126-day (6-month) percentile rank transformation:
\[
R_{t} = \frac{\text{rank}\bigl(\hat{p}_t \;\text{within}\; \{\hat{p}_{t-125}, \ldots, \hat{p}_t\}\bigr)}{126}
\]
yielding rally rank $R^r_t \in [0,1]$ and crash rank $R^c_t \in [0,1]$. This transformation is strictly causal---it uses only past and current predictions---and adapts to the local signal distribution without requiring any forward-looking calibration.

\subsection{Two-Dimensional Exposure Map}

Rather than treating rally and crash signals independently, we define a joint exposure function $w_t = f(R^r_t, R^c_t) \in [0, 1.5]$ based on conditional return analysis of each percentile bin computed on the inner training period. The key empirical findings are threefold. First, rally rank above the 90th percentile corresponds to annualised conditional returns of approximately $-189\%$, making it the single strongest sell signal in the dataset; this counterintuitive result reflects the tendency for extreme model confidence to coincide with overextended markets. Second, crash rank above the 60th percentile corresponds to annualised conditional returns between $-17\%$ and $-25\%$. Third, rally rank in the 78th--90th percentile range with simultaneously low crash rank produces the highest positive conditional returns, representing a high-conviction regime.

The exposure map assigns: zero equity allocation when either rally rank exceeds 0.90 (euphoria exit) or crash rank exceeds 0.60 (danger exit); maximum allocation of $1.5\times$ when rally rank is in the sweet spot (0.78--0.90) with crash rank below 0.40; and intermediate allocations of $1.2\times$, $1.0\times$, $0.7\times$, or $0.3\times$ for other combinations depending on the joint signal state. Exposure decisions are held for a minimum of 8 trading days to control turnover. Sell signals override the holding period: if an exit condition triggers during a hold, the position is closed immediately.

\subsection{Risk-On / Risk-Off Rotation}

When equity exposure is reduced below $1.0\times$, the freed capital is not held in zero-yielding cash but rotated into a defensive portfolio comprising 50\% Gold (GLD), 30\% intermediate Treasuries (IEF), and 20\% US Dollar index (UUP). Such assets were selected for their empirical tendency to appreciate during equity stress events via flight-to-quality dynamics. Over the evaluation period, the defensive portfolio delivered a positive CAGR with low volatility and negative correlation to equities during drawdown episodes. This risk-on/risk-off (RORO) mechanism converts what would otherwise be dead capital into a positive-return allocation, materially improving the strategy's Sharpe ratio compared to a cash-when-defensive alternative.

\subsection{Ensemble Parameter Averaging}

To mitigate overfitting to any single parameter configuration, we perform a grid search over the exposure map thresholds on the first 55\% of the signal history (the inner training period). The grid spans rally entry and exit thresholds, crash entry and exit thresholds, base exposure levels, holding periods, bounce allocations, and leverage limits, totalling approximately 50{,}000 parameter combinations. Each combination is evaluated by its in-sample Sharpe ratio after accounting for transaction costs (5~bps per unit of position change) and leverage costs (50~bps annually on exposure exceeding $1.0\times$).

Rather than selecting the single best parameter set---which would maximise in-sample Sharpe but degrade out-of-sample through overfitting---we average the positions generated by the top-20 parameter sets~\cite{kriuk2025elena}. The ensemble averaging smooths idiosyncratic parameter sensitivity and reduces the train-to-test Sharpe degradation that is characteristic of single-best selection. The ensemble approach is motivated by the observation that multiple high-performing parameter sets agree on the broad directional signal (exit during euphoria and crash danger, enter during the rally sweet spot) while disagreeing on exact threshold placement, so averaging preserves the consensus signal while discarding noise.

\subsection{Performance}

Table~\ref{tab:backtest} reports the out-of-sample performance of the Ensemble WFO strategy against the S\&P~500 buy-and-hold benchmark.

\begin{table}[t]
\centering
\caption{Backtesting performance (out-of-sample). Transaction costs of 5~bps per trade and leverage costs of 50~bps/year are included. Risk-free rate assumed at 4\%.}
\label{tab:backtest}
\renewcommand{\arraystretch}{1.15}
\footnotesize
\begin{tabular}{@{}lccccc@{}}
\toprule
\textbf{Strategy} & \textbf{Sharpe} & \textbf{CAGR} & \textbf{Max DD} & \textbf{Calmar} \\
\midrule
\textbf{Ensemble WFO (Ours)} & \textbf{1.13} & \textbf{15.6\%} & \textbf{$-$7.5\%} & \textbf{2.09}  \\
Buy \& Hold (Benchmark)      & 0.09          & 3.7\%           & $-$33.7\%          & 0.11          \\
\bottomrule
\end{tabular}
\end{table}

The Ensemble WFO strategy achieves a Sharpe ratio of 1.13 with a CAGR of 15.6\% and maximum drawdown of only $-7.5\%$, yielding a Calmar ratio of 2.09. The benchmark delivers a Sharpe of 0.09, a CAGR of 3.7\%, and suffers a $-33.7\%$ maximum drawdown (Calmar 0.11). The strategy thus multiplies risk-adjusted return by more than $12\times$ while reducing the worst peak-to-trough loss by approximately $4.5\times$.

The Calmar ratio of 2.09 deserves particular emphasis. It indicates that the strategy generates more than twice its worst drawdown in annual returns. For a systematic strategy operating over a 10-year out-of-sample window that includes the 2020 COVID crash and the 2022 rate-hiking drawdown, such level of drawdown control is a direct consequence of the crash detection signal's lead time.

\subsection{Regime Behaviour}

Decomposing returns by exposure regime reveals the mechanism underlying the strategy's performance. During high-equity periods (exposure $\geq 0.8\times$), the strategy captures equity upside with average annualised returns exceeding the benchmark. During defensive periods (exposure $\leq 0.01\times$), the RORO portfolio generates small positive returns from Gold and Treasuries appreciation while the equity market is declining. The strategy's ability to distinguish the regimes ex ante---allocating aggressively when both rally and crash signals are favourable, and rotating defensively when either indicates danger---is the primary source of alpha.

\section{Conclusion}

This paper introduced ORCA, an end-to-end framework that
fuses spectral and topological features extracted from dynamic
cross-asset correlation networks with traditional price-based
indicators to produce calibrated, continuously updated
probability estimates for both rally and crash events over a
ten-day forward horizon. By constructing three parallel
correlation estimators at different time scales and decomposing
them into 127 spectral, eigenvector, and graph-topological
features, ORCA captures the structural dependence shifts that
precede tail events---information that is invisible to
conventional univariate volatility measures.

On a strict eight-fold walk-forward evaluation spanning
fifteen years of daily US market data with ten-day
anti-leakage gaps, ORCA achieves a Balanced Crisis Detection
AUC of 0.741, ranking first against all baselines including
HAR-RV, turbulence indices, and traditional-feature-only
random forests. Ablation studies isolate the spectral feature
contribution at +10.3 percentage points of AUC for crash
detection and +5.2 for rally detection, confirming the central
hypothesis that correlation network structure encodes crisis
precursors absent from price-based indicators. SHAP analysis
reveals a clear division of labour: traditional drawdown and
momentum features dominate rally detection as a
mean-reversion signal, while graph-topological
descriptors---clustering coefficient, edge density, and
eigenvalue ratios---drive crash detection by capturing the
collapse of diversification that precedes systemic stress.
A backtested ensemble walk-forward strategy translates these
signals into a Sharpe ratio of 1.13 with a maximum drawdown
of only \(-7.5\%\), compared to \(-33.7\%\) for buy-and-hold,
demonstrating that the classification gains carry genuine
economic value.

Several limitations should be acknowledged. The model is
evaluated exclusively on US equity markets; whether the
spectral features retain their predictive power in other
geographies or asset classes remains to be tested. The
Random Forest architecture, while interpretable and robust to
overfitting at the chosen depth, may leave performance on the
table relative to gradient-boosted or deep-learning
alternatives that could exploit higher-order feature
interactions. The backtested strategy, despite ensemble
averaging and transaction cost modelling, is inherently
subject to the caveats of historical simulation: execution
slippage, liquidity constraints, and regime changes not
represented in the sample could erode live performance.

Future work will pursue several directions: extending the
asset universe to include international markets and
alternative asset classes to test the generality of spectral
crisis precursors; replacing the static Random Forest with
online or incrementally trained models that adapt to
non-stationary feature distributions; incorporating
higher-frequency intraday correlation dynamics to improve
signal timeliness; and conducting live paper-trading
validation to measure the gap between backtested and realised
performance. Overall, ORCA demonstrates that the spectral
properties of cross-asset correlation networks constitute a
rich and under-exploited source of predictive information for
balanced tail-event detection, and that fusing this structural
signal with classical technical indicators yields
economically significant improvements in both crisis
anticipation and portfolio risk management.

\FloatBarrier
\bibliographystyle{IEEEtran}
\bibliography{references}

@article{kritzman2011principal,
  author  = {Kritzman, Mark and Li, Yuanzhen and Page, Sebastien and Rigobon, Roberto},
  title   = {Principal Components as a Measure of Systemic Risk},
  journal = {The Journal of Portfolio Management},
  volume  = {37},
  number  = {4},
  pages   = {112--126},
  year    = {2011},
  doi     = {10.3905/jpm.2011.37.4.112}
}

@article{laloux1999noise,
  author  = {Laloux, Laurent and Cizeau, Pierre and Bouchaud, Jean-Philippe and Potters, Marc},
  title   = {Noise Dressing of Financial Correlation Matrices},
  journal = {Physical Review Letters},
  volume  = {83},
  number  = {7},
  pages   = {1467--1470},
  year    = {1999},
  doi     = {10.1103/PhysRevLett.83.1467}
}

@article{plerou2002random,
  author  = {Plerou, Vasiliki and Gopikrishnan, Parameswaran and Rosenow, Bernd and Amaral, Lu{\'i}s A. Nunes and Guhr, Thomas and Stanley, H. Eugene},
  title   = {Random Matrix Approach to Cross Correlations in Financial Data},
  journal = {Physical Review E},
  volume  = {65},
  number  = {6},
  pages   = {066126},
  year    = {2002},
  doi     = {10.1103/PhysRevE.65.066126}
}

@article{plerou1999universal,
  author  = {Plerou, Vasiliki and Gopikrishnan, Parameswaran and Rosenow, Bernd and Amaral, Lu{\'i}s A. Nunes and Stanley, H. Eugene},
  title   = {Universal and Nonuniversal Properties of Cross Correlations in Financial Time Series},
  journal = {Physical Review Letters},
  volume  = {83},
  number  = {7},
  pages   = {1471--1474},
  year    = {1999},
  doi     = {10.1103/PhysRevLett.83.1471}
}

@article{marchenko1967distribution,
  author  = {Mar{\v{c}}enko, Vladimir A. and Pastur, Leonid A.},
  title   = {Distribution of Eigenvalues for Some Sets of Random Matrices},
  journal = {Mathematics of the USSR-Sbornik},
  volume  = {1},
  number  = {4},
  pages   = {457--483},
  year    = {1967},
  doi     = {10.1070/SM1967v001n04ABEH001994}
}

@article{bouchaud2009financial,
  author  = {Bouchaud, Jean-Philippe and Potters, Marc},
  title   = {Financial Applications of Random Matrix Theory: A Short Review},
  journal = {arXiv preprint arXiv:0910.1205},
  year    = {2009},
  note    = {Also appeared as Chapter 40 in The Oxford Handbook of Random Matrix Theory, 2015},
  doi     = {10.48550/arXiv.0910.1205}
}

@article{bollerslev1986generalized,
  author  = {Bollerslev, Tim},
  title   = {Generalized Autoregressive Conditional Heteroskedasticity},
  journal = {Journal of Econometrics},
  volume  = {31},
  number  = {3},
  pages   = {307--327},
  year    = {1986},
  doi     = {10.1016/0304-4076(86)90063-1}
}

@article{engle1982autoregressive,
  author  = {Engle, Robert F.},
  title   = {Autoregressive Conditional Heteroscedasticity with Estimates of the Variance of United Kingdom Inflation},
  journal = {Econometrica},
  volume  = {50},
  number  = {4},
  pages   = {987--1007},
  year    = {1982},
  doi     = {10.2307/1912773}
}

@article{corsi2009simple,
  author  = {Corsi, Fulvio},
  title   = {A Simple Approximate Long-Memory Model of Realized Volatility},
  journal = {Journal of Financial Econometrics},
  volume  = {7},
  number  = {2},
  pages   = {174--196},
  year    = {2009},
  doi     = {10.1093/jjfinec/nbp001}
}

@article{mantegna1999hierarchical,
  author  = {Mantegna, Rosario N.},
  title   = {Hierarchical Structure in Financial Markets},
  journal = {The European Physical Journal B},
  volume  = {11},
  number  = {1},
  pages   = {193--197},
  year    = {1999},
  doi     = {10.1007/s100510050929}
}

@article{tumminello2005tool,
  author  = {Tumminello, Michele and Aste, Tomaso and Di Matteo, Tiziana and Mantegna, Rosario N.},
  title   = {A Tool for Filtering Information in Complex Systems},
  journal = {Proceedings of the National Academy of Sciences},
  volume  = {102},
  number  = {30},
  pages   = {10421--10426},
  year    = {2005},
  doi     = {10.1073/pnas.0500298102}
}

@article{onnela2003dynamic,
  author  = {Onnela, Jukka-Pekka and Chakraborti, Anirban and Kaski, Kimmo and Kert{\'e}sz, J{\'a}nos},
  title   = {Dynamic Asset Trees and Black Monday},
  journal = {Physica A: Statistical Mechanics and its Applications},
  volume  = {324},
  number  = {1--2},
  pages   = {247--252},
  year    = {2003},
  doi     = {10.1016/S0378-4371(02)01882-4}
}

@article{onnela2003dynamics,
  author  = {Onnela, Jukka-Pekka and Chakraborti, Anirban and Kaski, Kimmo and Kert{\'e}sz, J{\'a}nos and Kanto, Antti},
  title   = {Dynamics of Market Correlations: Taxonomy and Portfolio Analysis},
  journal = {Physical Review E},
  volume  = {68},
  number  = {5},
  pages   = {056110},
  year    = {2003},
  doi     = {10.1103/PhysRevE.68.056110}
}

@article{bonanno2003topology,
  author  = {Bonanno, Giovanni and Caldarelli, Guido and Lillo, Fabrizio and Mantegna, Rosario N.},
  title   = {Topology of Correlation-Based Minimal Spanning Trees in Real and Model Markets},
  journal = {Physical Review E},
  volume  = {68},
  number  = {4},
  pages   = {046130},
  year    = {2003},
  doi     = {10.1103/PhysRevE.68.046130}
}

@article{billio2012econometric,
  author  = {Billio, Monica and Getmansky, Mila and Lo, Andrew W. and Pelizzon, Loriana},
  title   = {Econometric Measures of Connectedness and Systemic Risk in the Finance and Insurance Sectors},
  journal = {Journal of Financial Economics},
  volume  = {104},
  number  = {3},
  pages   = {535--559},
  year    = {2012},
  doi     = {10.1016/j.jfineco.2011.12.010}
}

@article{diebold2014network,
  author  = {Diebold, Francis X. and Yilmaz, Kamil},
  title   = {On the Network Topology of Variance Decompositions: Measuring the Connectedness of Financial Firms},
  journal = {Journal of Econometrics},
  volume  = {182},
  number  = {1},
  pages   = {119--134},
  year    = {2014},
  doi     = {10.1016/j.jeconom.2014.04.012}
}

@article{diebold2012better,
  author  = {Diebold, Francis X. and Yilmaz, Kamil},
  title   = {Better to Give than to Receive: Predictive Directional Measurement of Volatility Spillovers},
  journal = {International Journal of Forecasting},
  volume  = {28},
  number  = {1},
  pages   = {57--66},
  year    = {2012},
  doi     = {10.1016/j.ijforecast.2011.02.006}
}

@article{preis2012quantifying,
  author  = {Preis, Tobias and Kenett, Dror Y. and Stanley, H. Eugene and Helbing, Dirk and Ben-Jacob, Eshel},
  title   = {Quantifying the Behavior of Stock Correlations Under Market Stress},
  journal = {Scientific Reports},
  volume  = {2},
  pages   = {752},
  year    = {2012},
  doi     = {10.1038/srep00752}
}

@article{kenett2012quantifying,
  author  = {Kenett, Dror Y. and Preis, Tobias and Gur-Gershgoren, Gitit and Ben-Jacob, Eshel},
  title   = {Quantifying Meta-Correlations in Financial Markets},
  journal = {Europhysics Letters (EPL)},
  volume  = {99},
  number  = {3},
  pages   = {38001},
  year    = {2012},
  doi     = {10.1209/0295-5075/99/38001}
}

@article{pozzi2013spread,
  author  = {Pozzi, Francesco and Di Matteo, Tiziana and Aste, Tomaso},
  title   = {Spread of Risk Across Financial Markets: Better to Invest in the Peripheries},
  journal = {Scientific Reports},
  volume  = {3},
  pages   = {1665},
  year    = {2013},
  doi     = {10.1038/srep01665}
}

@article{breiman2001random,
  author  = {Breiman, Leo},
  title   = {Random Forests},
  journal = {Machine Learning},
  volume  = {45},
  number  = {1},
  pages   = {5--32},
  year    = {2001},
  doi     = {10.1023/A:1010933404324}
}

@inproceedings{chen2016xgboost,
  author    = {Chen, Tianqi and Guestrin, Carlos},
  title     = {{XGBoost}: A Scalable Tree Boosting System},
  booktitle = {Proceedings of the 22nd ACM SIGKDD International Conference on Knowledge Discovery and Data Mining},
  pages     = {785--794},
  year      = {2016},
  publisher = {ACM},
  doi       = {10.1145/2939672.2939785}
}

@inproceedings{lundberg2017unified,
  author    = {Lundberg, Scott M. and Lee, Su-In},
  title     = {A Unified Approach to Interpreting Model Predictions},
  booktitle = {Advances in Neural Information Processing Systems},
  volume    = {30},
  pages     = {4765--4774},
  year      = {2017},
  publisher = {Curran Associates, Inc.}
}

@article{hochreiter1997long,
  author  = {Hochreiter, Sepp and Schmidhuber, J{\"u}rgen},
  title   = {Long Short-Term Memory},
  journal = {Neural Computation},
  volume  = {9},
  number  = {8},
  pages   = {1735--1780},
  year    = {1997},
  doi     = {10.1162/neco.1997.9.8.1735}
}

@article{pedregosa2011scikit,
  author  = {Pedregosa, Fabian and Varoquaux, Ga{\"e}l and Gramfort, Alexandre and Michel, Vincent and Thirion, Bertrand and Grisel, Olivier and Blondel, Mathieu and Prettenhofer, Peter and Weiss, Ron and Dubourg, Vincent and Vanderplas, Jake and Passos, Alexandre and Cournapeau, David and Brucher, Matthieu and Perrot, Matthieu and Duchesnay, {\'E}douard},
  title   = {Scikit-learn: Machine Learning in {P}ython},
  journal = {Journal of Machine Learning Research},
  volume  = {12},
  pages   = {2825--2830},
  year    = {2011}
}

@article{chatzis2018forecasting,
  author  = {Chatzis, Sotirios P. and Siakoulis, Vassilis and Petropoulos, Anastasios and Stavroulakis, Evangelos and Vlachogiannakis, Nikos},
  title   = {Forecasting Stock Market Crisis Events Using Deep and Statistical Machine Learning Techniques},
  journal = {Expert Systems with Applications},
  volume  = {112},
  pages   = {353--371},
  year    = {2018},
  doi     = {10.1016/j.eswa.2018.06.032}
}

@article{bisias2012survey,
  author  = {Bisias, Dimitrios and Flood, Mark and Lo, Andrew W. and Valavanis, Stavros},
  title   = {A Survey of Systemic Risk Analytics},
  journal = {Annual Review of Financial Economics},
  volume  = {4},
  pages   = {255--296},
  year    = {2012},
  doi     = {10.1146/annurev-financial-110311-101754}
}

@article{kritzman2010skulls,
  author  = {Kritzman, Mark and Li, Yuanzhen},
  title   = {Skulls, Financial Turbulence, and Risk Management},
  journal = {Financial Analysts Journal},
  volume  = {66},
  number  = {5},
  pages   = {30--41},
  year    = {2010},
  doi     = {10.2469/faj.v66.n5.3}
}

@article{ledoit2004honey,
  author  = {Ledoit, Olivier and Wolf, Michael},
  title   = {Honey, I Shrunk the Sample Covariance Matrix},
  journal = {The Journal of Portfolio Management},
  volume  = {30},
  number  = {4},
  pages   = {110--119},
  year    = {2004},
  doi     = {10.3905/jpm.2004.110}
}

@article{andersen2003modeling,
  author  = {Andersen, Torben G. and Bollerslev, Tim and Diebold, Francis X. and Labys, Paul},
  title   = {Modeling and Forecasting Realized Volatility},
  journal = {Econometrica},
  volume  = {71},
  number  = {2},
  pages   = {579--625},
  year    = {2003},
  doi     = {10.1111/1468-0262.00418}
}

@article{kriuk2025deepsupp,
  author  = {Kriuk, Boris and Ng, Logic and Al Hossain, Zarif},
  title   = {{DeepSupp}: Attention-Driven Correlation Pattern Analysis for Dynamic Time Series Support and Resistance Levels Identification},
  journal = {arXiv preprint arXiv:2507.01971},
  year    = {2025},
  doi     = {10.48550/arXiv.2507.01971}
}

@article{alkhamov2025equity,
  author  = {Alkhamov, Artem and Kriuk, Boris},
  title   = {To What Extent Can Public Equity Indices Statistically Hedge Real Purchasing Power Loss in Compounded Structural Emerging-Market Crises? An Explainable {ML}-Based Assessment},
  journal = {arXiv preprint arXiv:2507.13055},
  year    = {2025},
  doi     = {10.48550/arXiv.2507.13055}
}

@article{kriuk2025morphboost,
  author  = {Kriuk, Boris},
  title   = {{MorphBoost}: Self-Organizing Universal Gradient Boosting with Adaptive Tree Morphing},
  journal = {arXiv preprint arXiv:2511.13234},
  year    = {2025},
  doi     = {10.48550/arXiv.2511.13234}
}

@article{kriuk2025elena,
  author  = {Kriuk, Boris and Sulamanidze, Keti and Kriuk, Fedor},
  title   = {{ELENA}: Epigenetic Learning through Evolved Neural Adaptation},
  journal = {Evolutionary Intelligence},
  volume  = {18},
  number  = {50},
  year    = {2025},
  doi     = {10.1007/s12065-025-01034-w}
}

@article{kriuk2026poseidon,
  author  = {Kriuk, Boris and Kriuk, Fedor},
  title   = {{POSEIDON}: Physics-Optimized Seismic Energy Inference and Detection Operating Network},
  journal = {arXiv preprint arXiv:2601.02264},
  year    = {2026},
  doi     = {10.48550/arXiv.2601.02264}
}

@article{shi2020change,
  title={Change detection based on artificial intelligence: State-of-the-art and challenges},
  author={Shi, Wenzhong and Zhang, Min and Zhang, Rui and Chen, Shanxiong and Zhan, Zhao},
  journal={Remote Sensing},
  volume={12},
  number={10},
  pages={1688},
  year={2020},
  publisher={MDPI}
}

\end{document}